\documentclass[11pt]{article}
\usepackage{epsfig}
\usepackage{dsfont}

\begin{document}  
\thispagestyle{empty}
\begin{flushright}
hep-lat/0503004
\end{flushright} 
\vskip15mm
\begin{center}
{\huge
Remnant index theorem and low-lying \vskip2mm
eigenmodes for twisted mass fermions}
\vskip11mm
{\bf Christof Gattringer and Stefan Solbrig}
\vskip3mm
Institut f\"ur Theoretische Physik, Universit\"at
Regensburg \\
D-93040 Regensburg, Germany 
\vskip20mm
\begin{abstract}  
We analyze the low-lying spectrum and eigenmodes of lattice Dirac 
operators with a twisted mass term. The twist term expels the 
eigenvalues from a strip in the complex plane and all eigenmodes obtain
a non-vanishing matrix element with $\gamma_5$. For a twisted 
Ginsparg-Wilson operator the spectrum is located on two arcs
in the complex plane. Modes due to non-trivial topological charge  
of the underlying gauge field have their eigenvalues at the edges
of these arcs and obey a remnant index theorem. For configurations 
in the confined phase we find that the twist mainly affects the 
zero modes, while the bulk of the spectrum is essentially unchanged.
\end{abstract}
\vskip10mm
{\sl To appear in Physics Letters B.}
\vskip5mm
\end{center}

\newpage
\setcounter{page}{1}

\noindent
Lattice fermions with a twisted mass term
have gained a lot of attention recently \cite{twist1,twist2,
aokitwist,simulation}.
Particularly appealing for numerical simulations is the fact that a twisted 
mass term cures the notorious problem with exceptional configurations 
at a very low cost. This allows for simulations with
relatively small pion mass close to the physical point \cite{simulation}. 

For small pion masses the role of topological configurations of the gauge
field and the corresponding dynamics of low-lying modes of the Dirac 
operator become more important for QCD phenomenology. 
Of particular interest is the index theorem \cite{index} 
and its manifestation on the lattice. While for 
chiral lattice Dirac operators, obeying the Ginsparg-Wilson equation 
\cite{giwi82}, an exact index theorem holds \cite{giwiindex}, for
other formulations only an approximate realization of the index theorem
can be expected. For Wilson and staggered fermions this has been
studied numerically \cite{wilsonindex,staggerindex}.
For twisted mass fermions the interplay of topology and low-lying
Dirac eigenmodes has not yet been analyzed. In this letter we close this
gap, combining analytical arguments and numerical results. 
\vskip5mm
\noindent
The twisted mass Dirac operator ${\cal D}$ for two mass-degenerate 
flavors of fer\-mi\-ons, $u$ and $d$, has the form 
$$
{\cal D} \; = \; \big( D_0 \, + \, m \big) \mathds{1}_2 \; + \; 
i \,\mu \, \gamma_5 \, \tau_3 \; \;\; \; , \; \; \; \; \; \; 
\tau_3 \, = \, \mbox{diag}\,(1,-1) \, ,
$$
with $\mathds{1}_2$ and $\tau_3$ acting in flavor space. Alternatively one
can define the Dirac operators for $u$ and $d$ separately 
by $D_u = D(\mu)$ and $D_d = D(-\mu)$ with
\begin{equation}
D(\mu) \; = \; D_0 \, + \, m \, + \, i \, \mu \, \gamma_5 \; .
\label{dmudef}
\end{equation} 
$D_0$ is a standard lattice Dirac operator and for the numerical results
presented here is chosen to be Wilson's Dirac operator.
$\mu$ is the twisted mass parameter and $m$ the regular mass. 
In our notation we assume $\mu \geq 0$. In this work the mass 
parameter $m$ is adjusted such that for $\mu = 0$ massless quarks 
are described (``maximal twist''). We remark, however, that for the study of
eigenvalues and eigenvectors $m$ is an irrelevant parameter, since it 
only shifts the whole spectrum in the complex plane.

As remarked, QCD with a twisted mass term is particularly interesting for
lattice simulations since exceptional configurations are avoided
because
\begin{eqnarray}
&& \hspace{-5mm}
\det[D_u] \det[D_d] \, = \, \det[D(\mu)] \det[D(- \mu)]  \, = \,
\det[\gamma_5 D(\mu) \gamma_5] \det[D(- \mu)] \, =  
\nonumber
\\
&& \hspace{-5mm}
\det[ (D_0^\dagger \!+\!m \!+\! i \mu \gamma_5 )
( D_0 \!+ \!m \! - \! i \mu \gamma_5 )] \, = \,
\det[ (D_0 \!+\!  m)^\dagger (D_0 \!+\! m) + \mu^2 ] \; , 
\label{determinant}
\end{eqnarray} 
showing that the determinant for two flavors 
is strictly positive. For staggered fermions the regular mass term already 
excludes exceptional configurations. For Wilson fermions fluctuations of real 
eigenvalues can become large and only for relatively high values of the regular 
mass are exceptional configurations ruled out. When using the Wilson 
operator as input in the overlap projection, these fluctuations can also
make the evaluation of the square root problematic. 

\begin{figure}[t]
\centerline{\epsfig{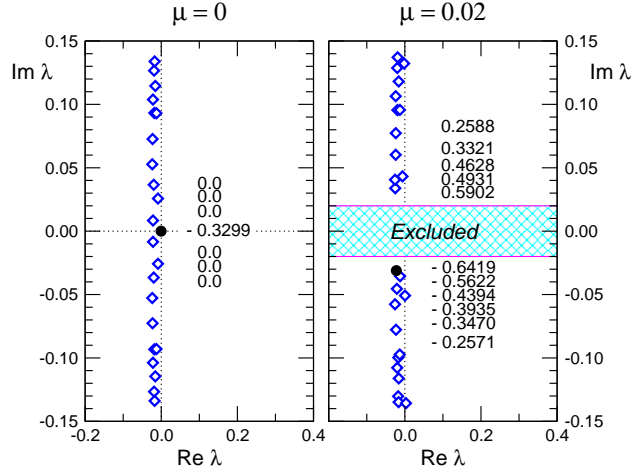}} 
\caption{Spectrum of the Wilson Dirac operator in the complex 
plane for vanishing twist (l.h.s.\ plot) and $\mu = 0.02$ 
(r.h.s.). For both plots the same quenched charge $Q = +1$ configuration
was used. The filled circle is used for the topological mode (zero-mode), 
while open diamonds represent bulk modes. 
The numbers next to the lowest eigenvalues
are the values of the $\gamma_5$ matrix elements (chirality)
for the corresponding eigenvectors.
\label{generalspec}}
\end{figure}

In the second line of Eq.\ (\ref{determinant}) we have used 
the $\gamma_5$-hermiticity of the Dirac operator $D_0$,
\begin{equation}
\gamma_5 \, D_0 \, \gamma_5 \; = \; D_0^\dagger \; .
\label{gamma5herm}
\end{equation}
In passing we note that for non-vanishing twist parameter $\mu$ this
equation is replaced by the generalized $\gamma_5$-hermiticity relation
\cite{generg5herm}
\begin{equation}
\gamma_5 \, D(\mu) \, \gamma_5 \; = \; D(- \mu)^\dagger \; .
\label{gamma5hermgen}
\end{equation}
Since the similarity transformation on the l.h.s.\ leaves the 
eigenvalues invariant, we conclude from (\ref{gamma5herm}) that the spectrum 
of $D(-\mu)$ is the complex conjugate of the spectrum of $D(\mu)$. This
property is reflected in the fact, that the product of 
determinants in (\ref{determinant}) is real. 
\vskip4mm
\noindent
Let us begin our analysis of the twisted mass Dirac operator $D(\mu)$ 
by comparing the spectrum with and without twist. In Fig.\ \ref{generalspec}
we show the low-lying eigenvalues for $\mu = 0$ (l.h.s.\ plot)
and $\mu = 0.02$ (r.h.s.) on the same
quenched configuration, generated with the L\"uscher-Weisz action
\cite{luweact} on a $16^4$ lattice at $\beta = 8.45$. The lattice spacing is
$a = 0.094$ fm as determined from the Sommer parameter in \cite{scale}. 
The eigenvalues $\lambda$ were computed with the implicitly restarted 
Arnoldi method \cite{arnoldi}. In the plots the numbers next to the symbols
representing the eigenvalues $\lambda$ are the chiralities of the corresponding
eigenvectors $\psi_\lambda$, i.e., the matrix elements
$(\psi_\lambda,\gamma_5 \psi_\lambda)$. The plots 
are representative for a larger ensemble of configurations we studied at 
several values of $\mu$, ranging from $\mu = 0.01$ to $\mu = 0.1$. 

We first discuss the spectrum for zero twist (l.h.s.\ plot).
The configuration we consider has topological charge $Q = +1$ and thus we
find a real eigenvalue near the origin. This ``topological eigenvalue''
is represented by a filled circle, while for the other eigenvalues
we use open diamonds. For a lattice Dirac operator 
obeying the Ginsparg-Wilson equation the topological 
eigenvalue would be exactly 
at the origin. For Wilson's Dirac operator $\gamma_5$-hermiticity
implies that real eigenvalues near the origin take over the
role of the zero-modes, since only for real eigenvalues $r$ the corresponding
eigenvector has non-trivial chirality, $(\psi_r, \gamma_5 \psi_r)
\neq 0$. This is evident from the $\gamma_5$ matrix elements
displayed next to the symbols for the eigenvalues. 

Upon turning on the twist parameter the spectrum changes considerably
(r.h.s.\ plot).
The eigenvalues are expelled from a strip along the real axis excluding
eigenvalues $\lambda$ with $| \mbox{Im} \lambda | < \mu$. Also 
the topological mode is shifted outside the ``forbidden strip.'' It is shifted 
downwards since, according to the index theorem, it has negative 
chirality for our $Q = +1$ configuration. The eigenvalues of topological 
modes with positive chirality are shifted upwards. This demonstrates 
that for $\mu \neq 0$ the spectrum is not symmetric with respect to
reflection at the real axis. We stress that this holds not only for the
topological eigenvalue, but also the other eigenvalues do not come in
complex conjugate pairs. For $\mu = 0$ the spectrum is symmetric as a 
consequence of (\ref{gamma5herm}). 
Another drastic change is the fact that 
for $\mu \neq 0$ all eigenvectors acquire non-vanishing chirality. They
have positive chirality for eigenvalues in the upper half and negative
chirality in the lower half (for positive $\mu$). The absolute value 
of the $\gamma_5$ matrix element decreases with increasing distance from the
real axis.
\vskip4mm
\noindent
Let us now try to understand these observations analytically. The 
$\gamma_5$ matrix element of an eigenvector $\psi_\lambda$
can be transformed as
\begin{eqnarray}
&& \hspace{-7mm}
\lambda (\psi_\lambda, \gamma_5 \psi_\lambda) = 
(\psi_\lambda, \gamma_5 D(\mu) \psi_\lambda) = 
(\psi_\lambda, D(-\mu)^\dagger \gamma_5 \psi_\lambda) = 
(D(-\mu) \psi_\lambda, \gamma_5 \psi_\lambda) = 
\nonumber 
\\
&& \hspace{-7mm}
([D(\mu)\! - \! i 2 \mu \gamma_5 ] \psi_\lambda, \gamma_5 \psi_\lambda) =
\lambda^* (\psi_\lambda, \gamma_5 \psi_\lambda) \! + \! i 2 \mu 
(\psi_\lambda, \psi_\lambda) = 
\lambda^* (\psi_\lambda, \gamma_5 \psi_\lambda) \! + \! i 2 \mu .
\nonumber 
\end{eqnarray}
In the second step we have used (\ref{gamma5hermgen}) and in the last step 
that the eigenvectors are normalized to 1. From this equation
follows
\begin{equation}
(\psi_\lambda, \gamma_5 \psi_\lambda) \; = \; \frac{\mu}{\mbox{Im} \,
\lambda} \; ,
\label{matrixelement}
\end{equation}
establishing that all eigenvectors have non-vanishing chirality which
decreases monotonically with $\mbox{Im} \, \lambda$. Using the fact
that $\gamma_5$ is bounded, i.e.,
$|(\psi_\lambda, \gamma_5 \psi_\lambda)| \leq 1$, we find ($\mu \geq 0$)
\begin{equation}
\mid \mbox{Im} \, \lambda \mid \; \geq \; \mu \; .
\label{imaginarybound}
\end{equation}
This result for the imaginary part is of course equivalent to the lower
bound for the determinant in Eq.\ (\ref{determinant}).

We remark that the fluctuations of the real part of the eigenvalues are
not reduced by the twisted mass term. This is obvious from comparing
the scatter of the eigenvalues in horizontal direction for the two plots
in Fig.~\ref{generalspec}, and was observed also on the larger sample of 
configurations we analyzed. For $\mu = 0$ the range of these fluctuations
is related to the Aoki phase. The fact that the horizontal spread of the
eigenvalues is essentially unchanged, indicates qualitatively that also for
twisted Wilson fermions  with small $\mu$ an Aoki phase has to be expected
(compare \cite{aokitwist}). 
\vskip4mm
\noindent
It is interesting to analyze the case of a twisted chiral operator, i.e.,
we now demand that the lattice Dirac operator $D_0$ obeys the Ginsparg-Wilson 
equation \cite{giwi82}
\begin{equation}
\gamma_5 D_0 \, + \, D_0 \gamma_5 \; = \; a D_0 \gamma_5 D_0 \, .
\label{giwi}
\end{equation}
Since a chiral $D_0$ has exact zero modes $\psi_0^\pm$ with 
definite chirality $\gamma_5 \psi_0^\pm = \pm \psi_0^\pm$, 
it is more suitable for analyzing the interplay 
between the twisted mass and topology. A twisted chiral Dirac operator 
provides a clean setting for such an analysis and the twisted Wilson
Dirac operator is expected to approximate the chiral behavior for sufficiently
smooth gauge fields.

It is straightforward, that adding a twisted mass term to 
a chiral $D_0$ turns $\psi_0^\pm$ into an eigenmode of $D(\mu)$
with eigenvalue $\lambda = \pm i \mu$.
However, more information can be extracted from the Ginsparg-Wilson
equation.
Using the definition (\ref{dmudef}) one has $D_0 = D(\mu) - i \mu \gamma_5$.
Inserting this expression into (\ref{giwi}) one finds after a few lines of
algebra (use (\ref{gamma5hermgen}) and $D(-\mu) = D(\mu) - i 2 \mu \gamma_5$), 
$$
D(\mu) \, + \, D(\mu)^\dagger \; = \; a D(\mu) D(\mu)^\dagger \, - \, a \mu^2
\; .
$$
Multiplying this equation with an eigenvector $\psi_\lambda$ from the right
and with $\psi_\lambda^\dagger$ from the left, it turns into an equation for
the eigenvalue $\lambda$, 
\begin{equation}
\lambda \, + \, \lambda^* \; = \; a \lambda \lambda^* \, - \, a \mu^2 \; .
\label{circle}
\end{equation}
Setting $\lambda = x + i y$ one finds that this is the equation
for a circle in the complex plane. This ``stretched Ginsparg-Wilson circle''
has center $1/a$ and radius $\sqrt{1/a^2 + \mu^2}$. Thus, in addition to 
obeying the
bound (\ref{imaginarybound}), the eigenvalues of a twisted chiral Dirac
operator are restricted to the stretched Ginsparg-Wilson circle. 
\begin{figure}[t]
\centerline{\epsfig{file=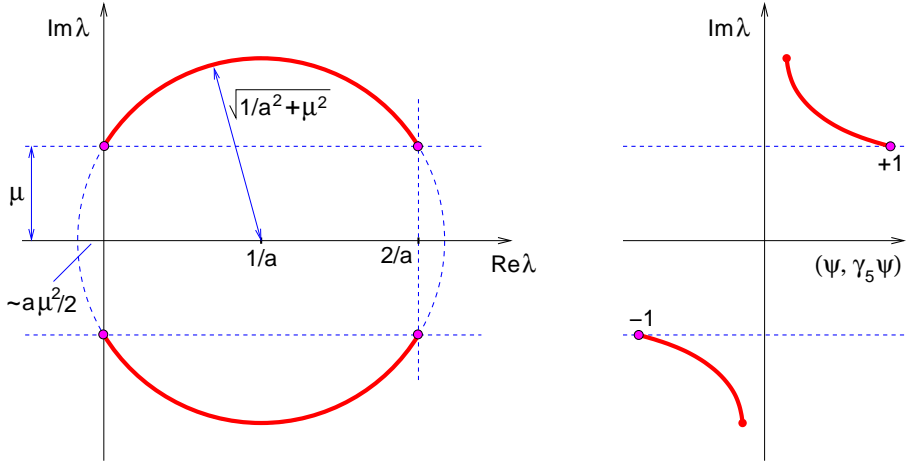,width=12cm,clip}} 
\caption{Graphical illustration of our results for spectrum 
(l.h.s.\ plot) and $\gamma_5$ matrix element (r.h.s.) for
twisted chiral fermions.  
\label{specillustration}}
\end{figure}

Our results for the spectrum of twisted chiral fermions are represented in the
l.h.s.\ plot of Fig.\ \ref{specillustration}. The eigenvalues are restricted
to the two arcs drawn as full curves. The zero-modes of the untwisted chiral
operator are mapped to the edges $\pm i \mu$
of the arcs (the doublers of the zero modes
end up at the other edges at $2/a \pm i \mu$). 
The index theorem for zero twist turns into a ``remnant index theorem'' 
for the edge modes, 
$$
Q \; = \; N_- \, - \, N_+ \; ,
$$
where $N_\pm$ is the number of eigenvalues $\lambda = \pm i \mu$. The 
corresponding eigenvectors have chiralities $\pm 1$.

The largest imaginary parts that can be reached by the eigenvalues restricted
to the arcs are $\mbox{Im} \, \lambda = \pm \sqrt{1/a^2 + \mu^2}$. 
According to (\ref{matrixelement}), this implies a lower bound for
the $\gamma_5$ matrix element, and we conclude
$$
\frac{\mu}{\sqrt{1/a^2 + \mu^2}} \; \leq \; 
|(\psi_\lambda, \gamma_5 \psi_\lambda)| \leq 1 \; .
$$
The upper bound is reached only by the topological edge-modes (and their
doubler partners). The relation (\ref{matrixelement}) 
between the chirality $(\psi, \gamma_5 \psi)$
and $\mbox{Im} \, \lambda$ is illustrated in the r.h.s.\ plot of Fig.\
\ref{specillustration}.
\vskip5mm
\noindent
Analyzing the case of a twisted chiral fermion, we have seen that topology
plays an important role for the spectrum also for finite $\mu$. In particular
the topological edge modes are protected by chiral symmetry and are 
unchanged when varying $\mu$. However,
Wilson fermions are not chiral and one expects that the connection between 
topology, low-lying spectrum and chirality of eigenmodes is only approximate.
An instance of this approximate realization is apparent in the fact that
the topological mode in the r.h.s.\ plot of Fig.\ \ref{generalspec} is
indeed the eigenvalue closest to the real axis. It is, however, not located at 
exactly $-i \mu$, as would be the case for a twisted chiral Dirac operator. 

In order to study the behavior of the edge modes we computed the spectrum
and eigenvectors of twisted Wilson fermions in the background of a 
discretized smooth instanton with radius $\rho = 4 a$ on a $16^4$ lattice
(see \cite{instanton} for details of the discretization). 
\begin{figure}[t]
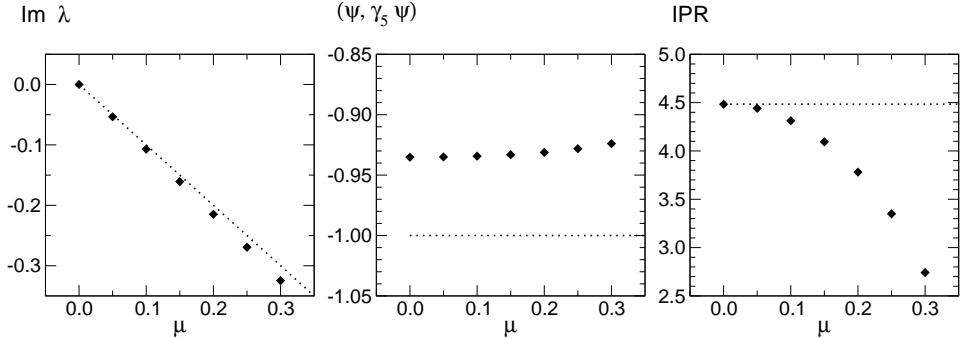

\hspace*{-1mm}
\centerline{
\epsfig{file=imlambda_vs_mu_16x16_R4.0.eps,height=4.5cm,clip}
\epsfig{file=g5_vs_mu_16x16_R4.0.eps,height=4.5cm,clip}
\epsfig{file=ipr_vs_mu_16x16_R4.0.eps,height=4.5cm,clip} 
}
\caption{Properties of the topological edge mode as a function of $\mu$.
We show the eigenvalue, the $\gamma_5$ matrix element and the
inverse participation ratio IPR (left to right). The dotted lines are 
the results for a twisted chiral Dirac operator, the symbols are
for twisted Wilson fermions.  
\label{edgemodeprops}}
\end{figure}
In Fig.\ \ref{edgemodeprops} we show, as a function of $\mu$,
the eigenvalue of the edge mode (l.h.s.\ plot), its chirality (center), 
and its inverse participation ratio (r.h.s.).
The inverse participation ratio is defined as $16^4 \sum_x (\psi^\dagger(x)
\psi(x))^2$. It is a measure for the localization of the eigenmode with
large values corresponding to localized modes while small values
indicate spread-out modes. In all three plots the symbols are the
numerical data, while the dotted lines are the analytical results for
twisted chiral fermions. The plots show a clear difference 
between the data and the expectation for chiral fermions. This indicates 
that twisted Wilson fermions do not give rise to a clear separation
of topological modes and bulk modes, which is only restored in the limit 
$\mu \rightarrow 0$. Repeating the same analysis with the chirally improved
Dirac operator \cite{chirimp}, we found that there the data points
\begin{figure}[t]
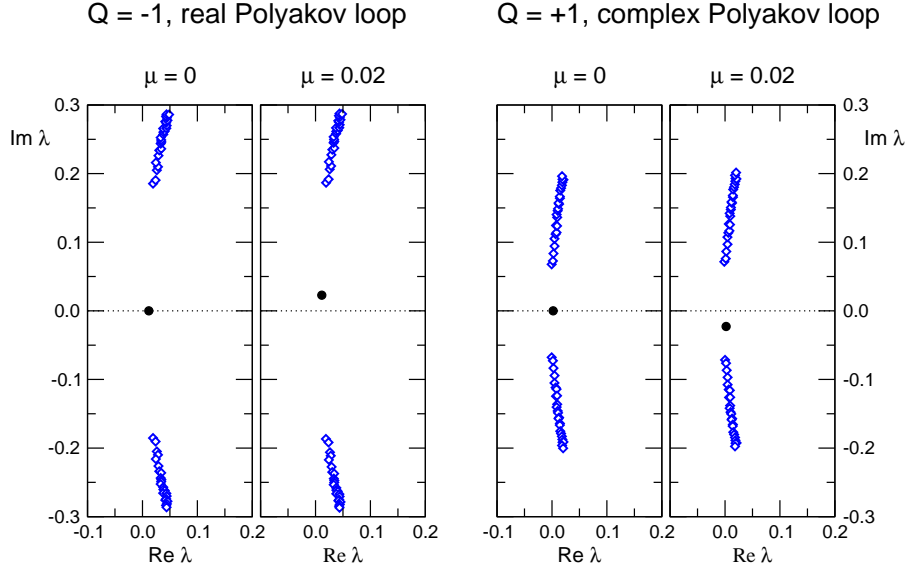

\centerline{\hspace*{-2mm}
\epsfig{file=sp_fintemp_realP.eps,height=7.5cm,clip} \hspace{3mm}
\epsfig{file=sp_fintemp_complexP.eps,height=7.5cm,clip}} 
\caption{Spectra at $\mu = 0, \, 0.02$ for configurations 
in the deconfined phase. 
\label{finitetemp}}
\end{figure}
fall on the chiral lines, demonstrating that for this case the topological 
edge mode is essentially protected by chiral symmetry. We remark, that since the 
chirally improved Dirac operator is only an approximate solution of the 
Ginsparg-Wilson equation, in this case the protection of the edge mode
is also only approximate. 
\vskip4mm
\noindent
A final question which we address in this letter is the role of the 
twisted mass term for configurations in the deconfined phase. There the 
spectrum acquires a gap near the origin and, according to the Banks-Casher 
formula \cite{baca}, the chiral condensate vanishes. 
In Fig.\ \ref{finitetemp} we show the effect of the twist term 
by comparing spectra with $\mu = 0$ and $\mu = 0.02$ for quenched 
configurations in the deconfined phase (L\"uscher Weisz action, $\beta = 8.45,
20^3 \times 6$). The l.h.s.\ pair of spectra is for a configuration 
with real Polyakov loop where a larger gap opens up \cite{bacastud}, 
while the right-hand side is for a configuration with complex Polyakov loop.
The plots show that in both cases, 
at least for moderately small $\mu$, the twisted mass
term affects only the topological mode, while the bulk of the spectrum is 
essentially unchanged. 

\newpage
\noindent
{\bf Acknowldegements:} 
We thank Andreas Sch\"afer for discussions and Christian Lang for 
interesting remarks on the relation
between the Dirac operator spectrum and the Aoki phase.
The calculations were done on the Hitachi SR8000
at the Leibniz Rechenzentrum in Munich and we thank the LRZ staff for
training and support. This work is supported by DFG and BMBF.

\end{document}